\documentclass[aps,prb,10pt,amsmath,showpacs,superscriptaddress,twocolumn]{revtex4-2} 
\usepackage[hmargin=2cm,vmargin=2.5cm]{geometry}

\usepackage{verbatim}

\usepackage[colorlinks,plainpages=false,linkcolor=blue,urlcolor=blue,citecolor=blue,pdfpagemode=UseNone,pdfstartview=FitBH]{hyperref} 
\usepackage[Charter,greekuppercase=italicized]{mathdesign} 
\usepackage{graphicx}    
\graphicspath{{./}{figs/}}
\DeclareGraphicsExtensions{.eps,.png,.pdf,.jpg}
\usepackage[version=4]{mhchem}

\newcommand{\sns}{Sn$_{0.15}$NbSe$_{1.75}$}
\newcommand{\snx}{Sn$_{x}$NbSe$_{2-\delta}$}

\newcommand{\di}{$dI/dV$}

\begin{document}

\title{Hybridization between surface flat bands and bulk bands in the topological nodal-line semimetal {\sns} probed via soft-point-contact spectroscopy}
\author{K.~Kumarasinghe}
\author{C.~Dissanayake}
\author{R.~Munir}
\author{M.~Tomlinson}
\affiliation{Department of Physics, University of Central Florida, Orlando, Florida 32816, USA}
\author{Y.~Nakajima}
\email[Corresponding author: ]{Yasuyuki.Nakajima@ucf.edu}
\affiliation{Department of Physics, University of Central Florida, Orlando, Florida 32816, USA}

\date{\today}

\begin{abstract}
  We report a detailed study of soft-point-contact spectroscopy of the superconducting topological nodal-line semimetal \sns\ with the superconducting transition temperature $T_{c}=9.5$ K. In the normal state, we observe prominent asymmetric double peaks in the differential conductance \di. The asymmetric \di\ curves are attributed to Fano resonance, quantum interference between two distinct tunneling paths of transmitting electrons into flat energy bands and dispersive bands. A phenomenological double Fano resonance model reveals the hybridization between these bands below the hybridization temperature $T_{\mathrm{hyb}}=23$ K. This hybridization drives an opening of a pseudogap below a characteristic temperature $T_{\mathrm{PG}}=6.8$ K. In the superconducting state, we observe an unusual upper critical field that increases linearly with decreasing temperatures from $0.4T_{c}$ to $0.01T_{c}$, suggestive of a possible exotic superconducting state. Our results suggest the presence of surface flat energy bands that stem from nontrivial topological nature of nodal lines in the bulk band structure and the hybridization between the surface flat bands and bulk bands in \sns.
\end{abstract}


\maketitle

\section{Introduction}

\begin{figure*}[tbh]
\includegraphics[width=0.95\linewidth]{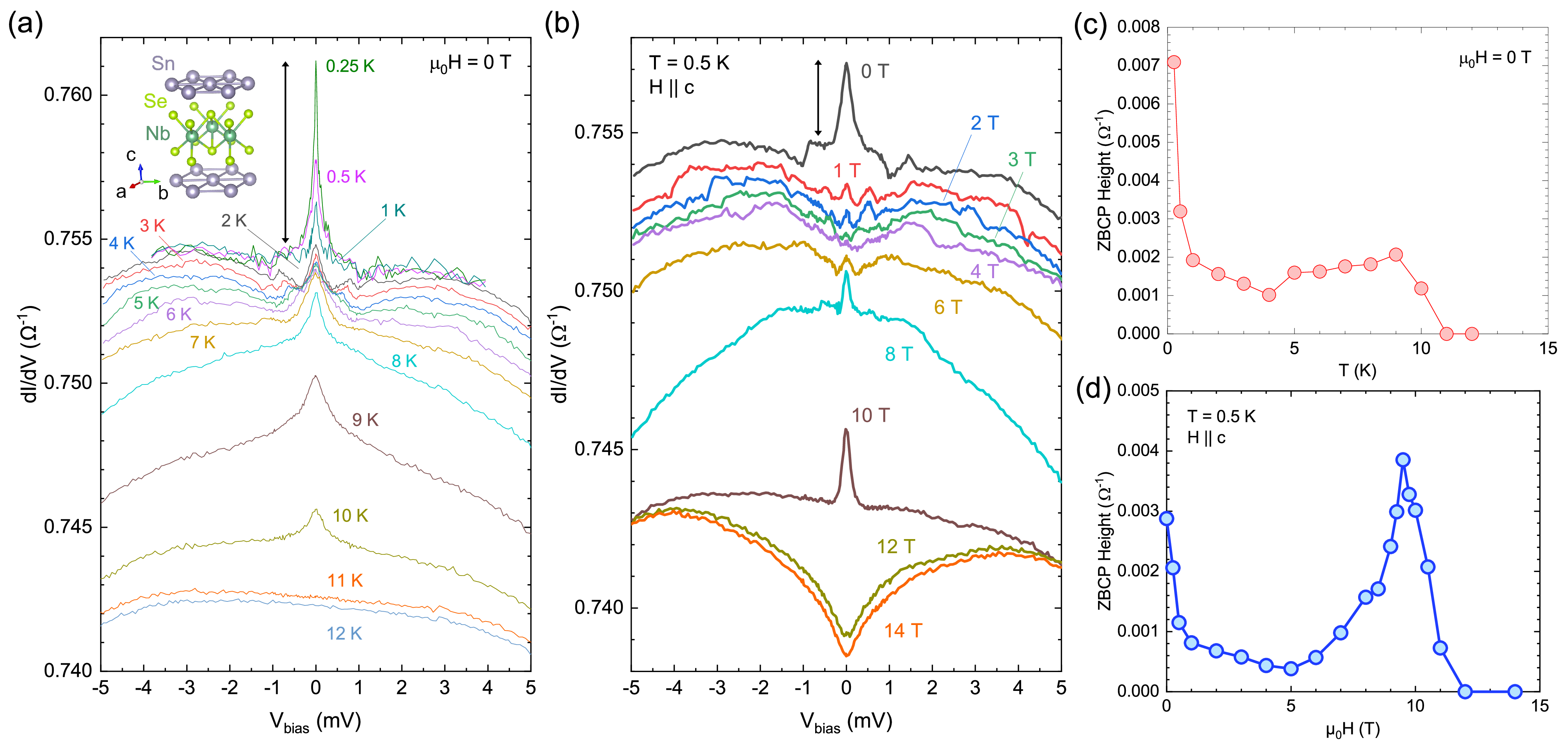}
\caption{\label{fig:ZBCPTH} (a) Differential conductance \di\ of \sns\ in zero magnetic field. Inset: noncentrosymmetric crystal structure of \ce{SnNbSe2}. (b) \di\ at 0.5 K in magnetic fields. Magnetic fields are applied parallel to the c axis. (c) ZBCP height as a function of temperatures in zero magnetic field. The ZBCP height is defined by the double arrow in the panel (a). (d) ZBCP height as a function of the magnetic field at 0.5 K. The ZBCP height is defined by the double arrow in the panel (b).}
\end{figure*}

Flat energy band systems represent one of the most exciting frontiers in the current condensed matter physics research. The dispersionless energy bands with the extremely large density of states near Fermi energy cause a plethora of exotic quantum states of matter, including ferromagnetism \cite{tasak98}, unconventional superconductivity \cite{cao18}, heavy fermion states \cite{stewa84}, fractional quantum Hall states \cite{tang11}, strange metallic states \cite{cao20}, and Mott insulating states \cite{cao18a}. These exotic states can be stabilized in various materials, such as magic-angle twisted bilayer graphene \cite{cao18,cao18a,cao20,das21,xie21}, moir\'e transition metal dichalcogenides \cite{park23,zeng23,cai23}, $f$-electron intermetallics \cite{stewa84}, and kagome materials \cite{li18b,yin19,kang20a,kang20b,ye24b}.

While these materials host flat energy bands in their bulk electronic structures, nondispersive energy bands can also originate from boundaries, i.e., surfaces on materials. A notable system that harbors flat surface energy bands is a topological nodal-line semimetal \cite{burko11,yu15,weng15a,huang16a}. In topological nodal-line semimetals, the bulk conduction and valence bands intersect in the Brillouin zone, forming closed loops instead of the discrete points found in Dirac and Weyl semimetals. These crossings are nonaccidental or symmetry-protected; thus, they cannot be lifted by perturbations without breaking the underlying symmetries. The topologically stable nodal lines in bulk bands give rise to approximately dispersionless surface states, so-called drumhead surface states, bounded by a projection of the bulk nodal line onto the surface plane. The dispersionless surface states provide a promising platform for the realization of high-temperature superconductivity \cite{kopni11} as well as topological superconducting states through bulk-surface coupling \cite{chang16,guan16}. Therefore, elucidating the effects of surface-bulk interaction is key to harnessing the topological nature, spin textures, and nondispersive energy bands of surface states in bulk topological nodal-line materials.

One of the promising materials for investigating the exotic phenomena arising from the coupling between nondispersive surface states and bulk states is \snx. This material shares a noncentrosymmetric crystal structure with \ce{ABSe2} (A = Sn, Pb and B = Nb, Ta) \cite{munir21,munir22} [the inset of FIG.\ref{fig:ZBCPTH}(a)]. The \ce{ABSe2} family is theoretically predicted to be a topological nodals line semimetal and exhibit superconductivity at low temperatures \cite{chen16a}. In fact, recent studies have experimentally revealed the drumhead surface states associated with topological nodal lines \cite{chang16,bian16,guan16} and superconductivity at 3.8 K \cite{ali14,wang15b,zhang16,chang16,sanka17} in \ce{PbTaSe2}. According to the theoretical calculations \cite{chen16a}, \ce{SnNbSe2} has bulk nodal lines located very close to Fermi energy and shows superconductivity at the highest superconducting transition temperature $T_{c}=7$ K. Interestingly, off-stoichiometric \snx\ shows unusual superconductivity. Its $T_{c}$ can be fine-tuned between 5 and 12 K--higher than the predicted value for \ce{SnNbSe2}--as Sn concentration $x$ and Se deficiency $\delta$ are controlled \cite{munir22}. Moreover, the in-plane upper critical field of low $T_{c}$ samples ($T_{c}\sim$ 5 K) exceeds the BCS Pauli paramagnetic limit by a factor of $\sim 2$, suggestive of a possible odd-parity superconducting pairing state \cite{munir21}. Despite these intriguing findings, the presence of the surface states in \sns remains unproven.

We here report the presence of flat-band surface states and the surface-bulk coupling in the tentative topological nodal-line semimetal \sns, probed via soft-point-contact spectroscopy. We observe asymmetric double peaks in differential conductance \di\ in the normal state, suggestive of Fano resonance, the quantum interference between tunneling paths of electrons into discrete and continuous energy states. Exploiting a phenomenological double Fano resonance model, we reveal the hybridization between the flat energy bands and bulk bands. We also observe an opening of a pseudogap in \di. The obtained hybridization gap amplitude $\Delta_{\mathrm{hyb}}$ is closely correlated with the pseudogap amplitude $\Delta_{\mathrm{PG}}$, as observed in the heavy fermion superconductor \ce{CeCoIn5} \cite{jang23}. In the superconducting state, we find a prominent zero-bias conductance peak (ZBCP) in \di\ and its unusual temperature and magnetic field dependence. The ZBCP most likely arises from an extrinsic origin. The upper critical field shows linear-in-$T$ behavior at low temperatures down to $0.01T_{c}$, different from that predicted by the Werthamer-Helfand-Hohenberg (WHH) theory for conventional type-II superconductors. The anomalus temperature dependence of $H_{c2}$ may be suggestive of an unusual superconducting state. Our findings suggest that the presence of surface flat bands, induced by topological nodal lines in the bulk bands, and the hybridization between the surface flat bands and bulk bands in \sns.

\section{Experimental Methods}

\sns\ single crystals were grown using a self-flux method \cite{munir21,munir22}. The excess of molten Sn flux was removed by centrifuging. We obtained a single-crystal x-ray diffraction pattern consistent with the noncentrosymmetric crystal structure of SnNbSe$_{2}$ with the space group $P\bar{6}m2$ and clearly different from the $2H$–\ce{NbSe2}-type structure of Sn-intercalated \ce{NbSe2} with Sn concentration up to 0.04. The atomic ratio of the crystal was determined with x-ray fluorescence spectroscopy. Soft-point-contact spectra were obtained by a lock-in technique with ac current modulation \cite{daghe10a,das19a}. An Au/Ag/\sns\ point-contact junction was fabricated using Ag paste on the $ab$ plane of the sample surface. We used a quasi-four-probe configuration to measure the point-contact spectra.

\section{Results and Discussion}

In the differential conductance \di\ of {\sns}, we observe anomalous behavior in both the normal and superconducting states. We plot \di\ of {\sns} as a function of bias voltage $V_{\mathrm{bias}}$ in zero magnetic field in FIG.\ref{fig:ZBCPTH}(a). In the normal state at high temperatures above 11 K, \di\ curves exhibit asymmetric parabola-like shapes. Upon cooling the temperature below 10 K, a peak appears at $V_{\mathrm{bias}}=0$, associated with superconductivity. The zero-bias-conductance peak (ZBCP) becomes more prominent and sharper at low temperatures, as manifested in the unusual temperature dependence of ZBCP height shown in FIG.\ref{fig:ZBCPTH}(c).

Superconductivity of \sns\ is suppressed by magnetic fields, leading to the emergence of double peaks in the normal state \di. \di\ of {\sns} at 0.5 K in applied magnetic fields is plotted as a function of $V_{\mathrm{bias}}$ in FIG.\ref{fig:ZBCPTH}(b). The magnetic fields are applied perpendicular to the ab plane. As the applied magnetic field increases, the ZBCP height is suppressed gradually until reaching $\mu_{0}H=5$ T. With increasing applied magnetic field above 5 T, the ZBCP is enhanced again, becoming particularly pronounce around 10 T in the vicinity of $H_{c2}$. The anomalous field dependence of the ZBCP is more clearly shown in FIG.\ref{fig:ZBCPTH}(d). Above $\mu_{0}H=$ 12 T, the ZBCP disappears, indicating that the superconductivity is completely suppressed. In the normal state above 12 T, notable double peaks around $\pm$4 mV are observed in \di\, accompanied by a sharp dip at $V_{\mathrm{bias}}=0$.

\begin{figure}[t]
\includegraphics[width=0.95\linewidth]{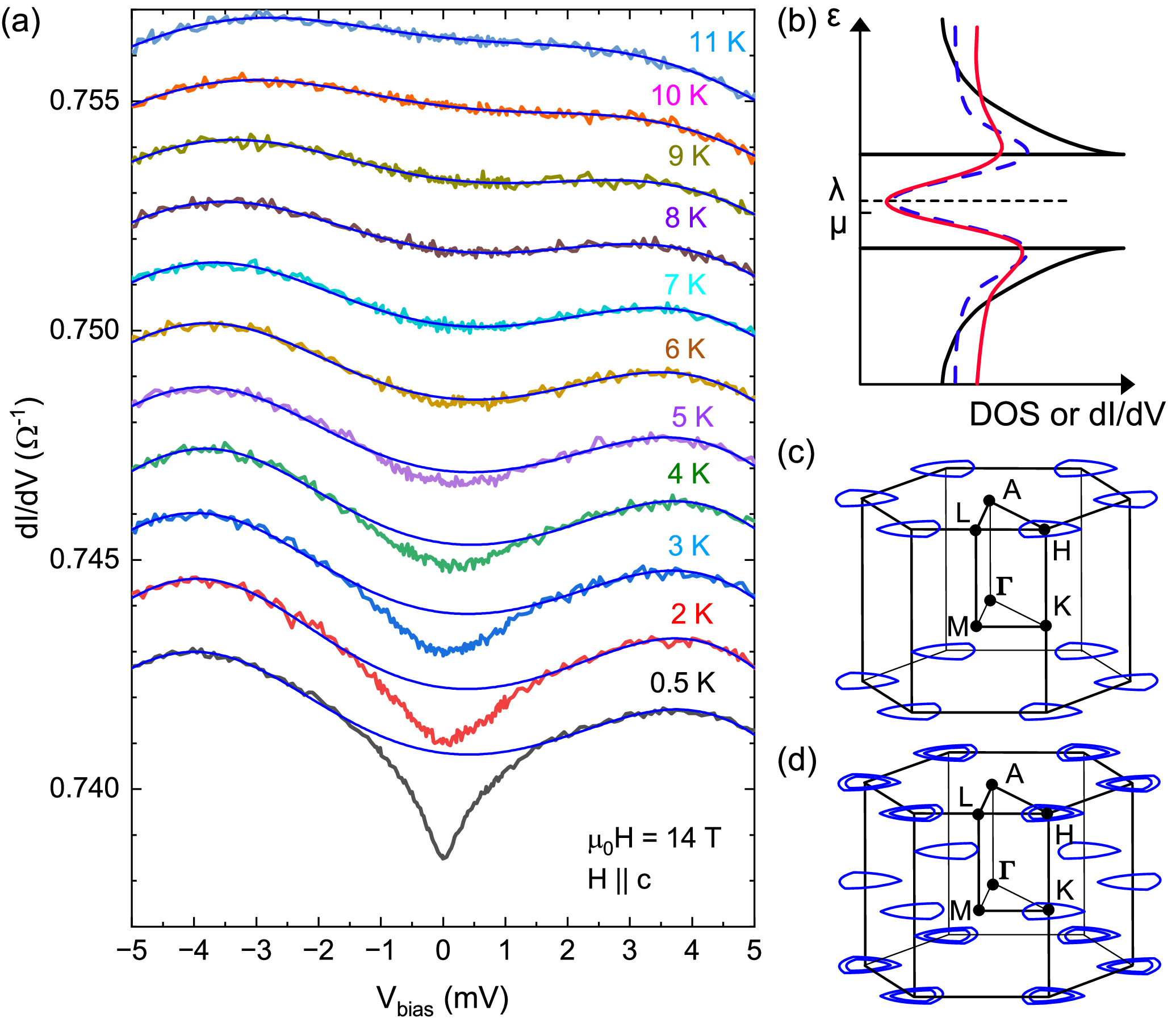}
\caption{\label{fig:dIdV14T} \di\ of {\sns} in an applied magnetic field of 14 T. The magnetic field is applied along the $c$ axis. For clarity, the \di\ curves are successively shifted upward by 0.0015 $\Omega^{-1}$. The blue solid lines are fits to the data using the double Fano resonance model, Eq.(\ref{eq:DFR}). (b) Schematic diagrams of density of states (DOS) for renormalized bands due to hybridization (thick line) and DOS broadened by correlation effects (blue dashed line). The red line is the expected \di\ spectrum with Fano resonance. $\mu$ is the chemical potential, and $\lambda$ is the resonance energy. (c) Schematic nodal lines predicted for \ce{ABSe2} without SOC and (d) with SOC.}
\end{figure}

The asymmetry in \di\ persists even at high magnetic fields, demonstrating its robustness against applied magnetic fields. In contrast, the double peaks in \di\ diminish as the temperature increases. Figure \ref{fig:dIdV14T} illustrates the evolution of the double peaks in \di\ at an applied magnetic field of $\mu_{0}H$ = 14 T. While the double peaks are prominent at 0.5 K, its amplitudes gradually decrease as the temperature increases.

\begin{figure}[tbh]              
\includegraphics[width=0.95\linewidth]{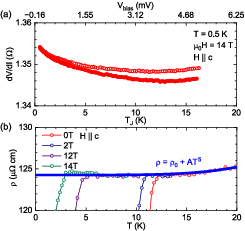}
\caption{\label{fig:TJ} (a) Differential resistance dV/dI as a function of $T_{J}=\sqrt{T_0^2 + V_{\mathrm{bias}}^2/4L_{0}}$. The corresponding $V_{\mathrm{bias}}$ is shown on the top axis. (b) Temperature dependence of the resistivity for \sns\. The blue solid line is a fit to data using $\rho=\rho_{0}+AT^{5}$, suggestive of dominant electron-phonon scattering over electron-electron scattering in the measured temperature range.}
\end{figure}

\begin{figure*}[tbh]
\includegraphics[width=0.95\linewidth]{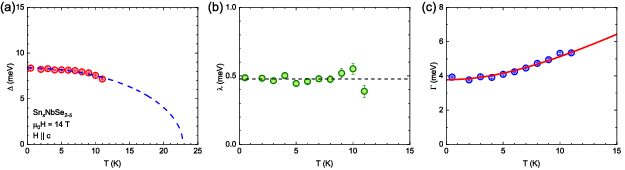}
\caption{\label{fig:GlD}  (a) Temperature dependence of the peak separation $\Delta$. The blue dashed line is a fit to the data using $\Delta(T) = \Delta(0)\sqrt{1-(T/T_{\mathrm{hyb}})^{2}}$ with $\Delta(0)=8.37 \pm 0.04$ meV and $T_{\mathrm{hyb}}=23\pm 1$ K. (b) The resonance energy $\lambda$ as a function of temperature. The black dashed line is the average value of the resonance energy over the measured $T$ range, $\bar{\lambda}=0.48$ meV, suggesting that $\lambda$ is independent of temperature. (c) Temperature dependence of the broadening parameter $\Gamma$ of {\sns} obtained from a fitting to Eq.~(\ref{eq:DFR}) at 14 T. The red solid line is a fit to the data using $\Gamma(T)=\sqrt{(\alpha k_{\mathrm{B}} T)^{2}+2(k_{\mathrm{B}}T_{\mathrm{K}})^{2}}$ with $\alpha = 4.0 \pm 0.1$ and $T_{\mathrm{K}}=31.0\pm 0.4$ K.}
\end{figure*}

We observe two distinct anomalous features in \di\ of \sns: 1) the asymmetric double peaks in the normal state and 2) ZBCP in the superconducting state. First, we will discuss the origin of the asymmetric \di\ in the normal state.

The asymmetric double peaks in \di\ of {\sns} are {\it not} caused by extrinsic nonspectroscopic effect, local heating at a point-contact junction in the thermal regime, where the mean free path $\ell$ is much shorter than the junction diameter $d$, $\ell \ll d$ \cite{naidy05}. In the thermal regime, the voltage drop occurs within the junction due to inelastic scattering, yielding local Joule heating. In this situation, the temperature at the point-contact junction, $T_{J}$, is given by,
\begin{equation}
  T_{J}^{2} = T_0^2 + V_{\mathrm{bias}}^2/4L_{0},\label{eq:TJ}
\end{equation}
where $T_{0}$ is the bath temperature and $L_{0} = (\pi e k_{\mathrm{B}})^{2}/3$ is the Lorenz number. Here, we assume that the Wiedemann-Franz law is valid and the Lorenz number is independent of temperature within the measured temperature range. If the junction is in the thermal regime, differential resistance $dV/dI$ as a function of $T_{J}$ mimics the temperature dependence of the bulk resistivity. To verify this, we plot $dV/dI$ at 0.5 K in an applied field of 14 T as a function of $T_{J}$ obtained from Eq. (\ref{eq:TJ}) in FIG.\ref{fig:TJ}. The upper curve (open symbols) is derived from the positive bias voltage, while the lower one (filled symbols) is from the negative bias voltage. Both $dV/dI$ curves show a clear upturn at low $T_{J}$ below $\sim$ 10 K. The upturn behavior observed in both $dV/dI$ is inconsistent with the temperature dependence of bulk resistivity, which exhibits $\rho\propto T^{5}$, as shown in FIG.\ref{fig:TJ}(b). This indicates that the junction of our sample is not in the thermal regime and the asymmetric \di\ results from an intrinsic origin.

Asymmetric \di\ can be attributed to Fano resonance, which has been observed in various systems with a localized flat energy band, including heavy fermion compounds \cite{park05,goll06,park12,zhang13,jaggi17,shiga21,yin21,shiga23} and kagome magnets \cite{das19a}. The Fano resonance arises from the quantum interference between two distinct tunneling paths: one tunneling into the localized band and the other into the conduction band. These two paths cause the Fano line shape in \di\ with a peak centered at the resonance energy, $\lambda$, which can be described by the Fano resonance model:
\begin{equation}
   G_{\mathrm{FR}}(\varepsilon) = \frac{|q-\varepsilon|^{2}}{1+\varepsilon^{2}},
\end{equation}
where $\varepsilon=(eV-\lambda)/\Gamma$, $\Gamma$ is a broadening parameter, and $q$ is a Fano parameter \cite{fano61}.

However, the Fano line shape alone cannot explain the observed double peaks in \di. Similar asymmetric \di\ with double peaks has indeed been observed in heavy fermion compounds, where Fano resonance is caused by the interference of the tunneling channels into itinerant conduction electrons and localized $f$ electrons \cite{park05,goll06,park12,zhang13,jaggi17,shiga21,yin21,shiga23}. In these compounds, the hybridization between the conduction and $f$ electrons induces a hybridization gap. This hybridization gap results in the double peaks and a dip at $\lambda$ in \di, as illustrated in FIG.\ref{fig:dIdV14T}(b). In fact, theoretical models that take into account the hybridization between conduction and $f$ electrons, proposed by Malteva, Dzero, and Coleman (the MDC model) \cite{malts09} and by Yang (the modified Fano model) \cite{yang09b}, successfully reproduce experimental observations in point-contact spectra of heavy fermion system \cite{park05,goll06,park12,zhang13,jaggi17,shiga21,yin21,shiga23}.

Although Fano resonance requires a localized energy state to interfere with continuous energy states, there are no bulk localized electrons in \sns, unlike heavy fermions or kagome materials. Instead, topological nodal-line semimetals, including  \ce{ABSe2}, can host dispersionless drumhead surface states associated with nodal lines \cite{burko11}. In the case of \ce{ABSe2}, nodal lines are located around the $H$ points without spin-orbit coupling (SOC) and additionally around the $K$ points with SOC \cite{chen16a}, as shown in FIGs.\ref{fig:dIdV14T} (c) and (d). Enclosed by the projections of these nodal lines onto the surface plane, dispersionless drumhead surface states emerge. While theoretical predictions to off-stoichiometric \sns\ are currently lacking, we propose that \sns\ is a topological nodal-line semimetal from analogy to the stoichiometric \ce{SnNbSe2}. In this scenario, we attribute the observed asymmetric double peaks in \di\ to the interference between two channels into the surface flat bands and bulk bands and the hybridization between these bands. 

Since the MDC model and the modified Fano model are primarily proposed for heavy-fermion systems, these models may not be applicable to Fano resonance due to the surface flat bands and the bulk bands in topological nodal-line semimetals. Thus, we fit the data to a phenomenological model that incorporates double Fano lines \cite{shiga21},
\begin{equation}
  G_{\mathrm{DFR}}(V) = s [G_{\mathrm{FR}}(\varepsilon_{+})+G_{\mathrm{FR}}(\varepsilon_{-})] + G_0, \label{eq:DFR}
\end{equation}
where $\varepsilon_{\pm} = (eV-\lambda\pm\Delta/2)/\Gamma$, $\Delta$ is the peak separation, $s$ is a scaling factor, and $G_{0}$ is a constant background conductance independent of $V_{\mathrm{bias}}$. $G_{0}=0.7348$ is fixed since it remains temperature-independent at 14 T. The obtained parameters $\Delta$, $\lambda$, and $\Gamma$ are plotted in FIG.\ref{fig:GlD}. As shown in FIG.\ref{fig:dIdV14T}(a), the theoretical curves are in excellent agreement with the experimental \di\ curves above 7 K. This result reveals two key features in this system: 1) the presence of topological surface flat bands induced by bulk nodal lines and 2) the hybridization between the surface flat bands and bulk bands.

The peak separation $\Delta$ is closely linked to the hybridization gap $\Delta_{\mathrm{hyb}}$ obtained, as reported in the point contact spectroscopy studies of \ce{EuNi2P2} \cite{shiga21} and \ce{UPd2Al3} \cite{jaggi17}. Temperature dependence of the obtained peak separation, $\Delta (T)$, is plotted in FIG.\ref{fig:GlD} (a). As the temperature increases, the amplitude of $\Delta(T)$ is gradually suppressed. To determine the characteristic temperature $T_{\mathrm{hyb}}$, at which the hybridization gap closes, we fit the data to an empirical expression, $\Delta (T)=\Delta(0)\sqrt{1-(T/T_{\mathrm{hyb}})^{2}}$. From this fitting, we obtain $\Delta(0)=8.37 \pm 0.04$ meV and $T_{\mathrm{hyb}}=23\pm 1$ K. According to the previously reported work on the point contact spectroscopy of \ce{EuNi2P2}, $\Delta$ shows similar temperature dependence to $\Delta_{\mathrm{hyb}}$ obtained from the MDC model and is scaled by a scaling factor $\eta=\Delta/\Delta_{\mathrm{hyb}} = 1.64\pm 0.03$ \cite{shiga21}. Utilizing this empirical scaling factor $\eta$, we obtain $\Delta_{\mathrm{hyb}}(0) = 5.3 \pm 0.1$ meV for \sns.

The obtained values of $\lambda$ and $\Gamma$ provide insights into the surface flat energy bands. As shown in FIG.\ref{fig:GlD}(b), $\lambda$ for \sns\ is $\sim$ 0.48 meV and nearly independent of temperature, indicating that the surface flat energy bands are located close to the Fermi energy. The broadening parameter $\Gamma$ increases with temperatures, varying as $\Gamma(T)=\sqrt{(\alpha k_{\mathrm{B}} T)^{2}+2(k_{\mathrm{B}}T_{\mathrm{K}})^{2}}$, the expression describing a single-site Kondo resonance \cite{nagao02}. From the theoretical curve, we extract $\alpha = 4.0 \pm 0.1$ and $T_{\mathrm{K}}=31.0\pm 0.4$ K. The obtained value of $\alpha$ is slightly larger than $\alpha =\pi$ for the single-site Kondo model. $T_{\mathrm{K}}=31$ K is also slightly larger than the gap-closing temperature $T_{\mathrm{hyb}}$. 

\begin{figure}[t]
\includegraphics[width=0.95\linewidth]{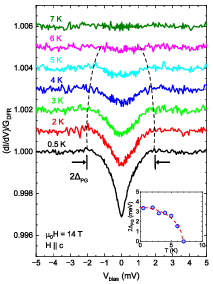}
\caption{\label{fig:PG} Normalized differential conductance $(dI/dV)/G_{\mathrm{DFR}}$ of {\sns} in an applied magnetic field of 14 T. The magnetic field is applied parallel to the $c$ axis. The normalized curves are successively shifted upward by 0.001 for clarity. The pseudogap amplitude $2\Delta_{\mathrm{PG}}$ is defined by the pseudogap width indicated by the arrows. Dashed lines are guides to the eyes. Inset: temperature dependene of $2\Delta_{\mathrm{PG}}$. The red dashed line is a fit to the data using $2\Delta_{\mathrm{PG}}(T) = 2\Delta_{\mathrm{PG}}(0)\sqrt{1-(T/T_{\mathrm{PG}})^{2}}$ with $2\Delta_{\mathrm{PG}}(0) = 3.4\pm0.1~\mathrm{meV}$.}
\end{figure}

The phenomenological double Fano resonance model excellently explains our \di\ curves above 7 K at 14 T. This finding suggests the presence of surface flat bands and the hybridization between the surface flat bands and bulk bands. However, at temperatures below 7 K, the measured \di\ curves show noticeable deviations from the theoretical curves around zero-bias voltage, suggestive of an opening of a pseudogap. To highlight the pseudogap-like dip, we normalize the measured \di\ curves by $G_{\mathrm{DFR}}(V)$ after fitting the background curves to the double Fano resonance model. The normalized curves are plotted in the FIG.\ref{fig:PG}. 

\begin{figure}[t]
\includegraphics[width=0.95\linewidth]{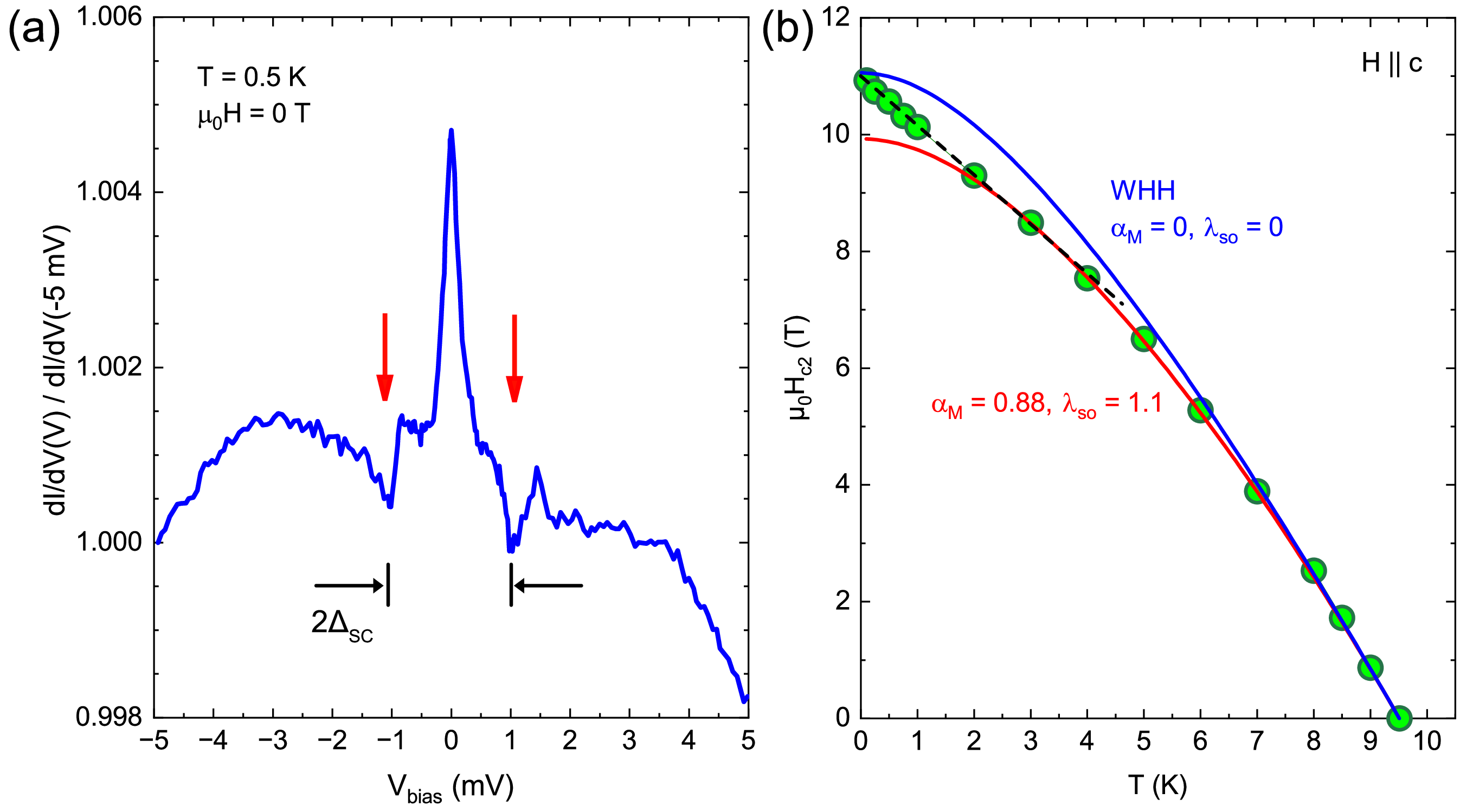}
\caption{\label{fig:dIdVnor0p5K} (a) Normalized differential conductance $(dI/dV)_{n}= [dI/dV(V)]/ [dI/dV (-5~\mathrm{mV})]$ of {\sns} at 0.5 K in zero magnetic field. The red arrows indicate dip features associated with the superconducting gap amplitude $2\Delta_{\mathrm{SC}}$. (b) Temperature dependence of upper critical field along the $c$ axis, obtained from the midpoint of resistive transitions in the point-contact junction at $V_{\mathrm{bias}}=0$. The blue solid line is a theoretical curve based on the WHH theory with $\alpha_{\mathrm{M}}=0$ and $\lambda_{\mathrm{SO}}=0$ and the red solid line is a theoretical curve with $\alpha_{\mathrm{M}}=0.88$ and $\lambda_{\mathrm{SO}}=1.1$. The black dashed line is a guide to the eyes.}
\end{figure}

The pseudogap of \sns\ is correlated with the hybridization gap. We define the pseudogap amplitude $2\Delta_{\mathrm{PG}}$ by the deviation from $(dI/dV)/G_{\mathrm{DFR}}=1$, as indicated by arrows in FIG.\ref{fig:PG}(a). The temperature dependence of $2\Delta_{\mathrm{PG}}$ is shown in the inset of FIG.\ref{fig:PG}. The red dashed line represents an empirical curve of a gap evolution, $2\Delta_{\mathrm{PG}}(T)=2\Delta_{\mathrm{PG}}(0)\sqrt{1-(T/T_{\mathrm{PG}})^{2}}$ with $2\Delta_{\mathrm{PG}}=3.4\pm 0.1$ meV and $T_{\mathrm{PG}}=6.8\pm 0.2$ K. The ratio of the pseudogap amplitude to the pseudogap temperature is $2\Delta_{\mathrm{PG}}(0)/k_{\mathrm{B}}T_{\mathrm{PG}} = 5.8 \pm 0.2$, which is in good agreement with $2\Delta_{\mathrm{hyb}}(0)/k_{\mathrm{B}}T_{\mathrm{hyb}}=5.4\pm 0.3$. This agreement implies that the hybridization gap and pseudogap are driven by the same mechanism, namely, the hybridization between flat surface bands and bulk bands. In fact, a hybridization-controlled pseudogap state is reported in \ce{CeCoIn5} \cite{jang23}. Our observation may suggest that the hybridization-driven pseudogap opening is a universal feature in systems with flat energy bands and dispersive energy bands.

We now examine \di\ in the superconducting state of \sns. Figure \ref{fig:dIdVnor0p5K}(a) shows normalized differential conductance $(dI/dV)_{n}= [dI/dV(V)]/ [dI/dV (-5~\mathrm{mV})]$ in zero magnetic field at 0.5 K. We observe clear dips at $\pm 1$ meV in (\di)$_{n}$, associated with the superconducting gap amplitude $2\Delta_{\mathrm{SC}}$. From this characteristic energy scale, we obtain $2\Delta_{\mathrm{SC}}/k_{\mathrm{B}}T_{c}=2.58$. We estimate $T_{c}$ = 9.5 K from the midpoint of the resistive transitions of the point-contact junction at $V_{\mathrm{bias}}=0$. This value is much smaller than the predicted value of 3.43 for BCS weak-coupling superconductors. This small value of $2\Delta_{\mathrm{SC}}/k_{\mathrm{B}}T_{c}$ can be ascribed to the multigap superconductivity. While the observed dips are associated with a smaller superconducting gap in a passive band, anomaly associated with a larger superconducting gap in a active band is possibly concealed by the background conductance arising from the double peaks and pseudogap due to the hybridization of the surface flat bands and bulk bands.

The upper critical field of \sns\ shows anomalous temperature dependence at low $T$. The upper critical field $H_{c2}$ parallel to the c axis is plotted in FIG.\ref{fig:dIdVnor0p5K}(b), along with theoretical curves calculated using the WHH model. We observe an apparent deviation from a WHH curve with the Maki parameter $\alpha_{\mathrm{M}}=0$ and the spin-orbit scattering parameter $\lambda_{\mathrm{SO}}=0$ at low temperatures below $6.5~\mathrm{K} \sim 0.7 T_{c}$, suggestive of the suppression of $H_{c2}$ due to paramagnetic effect. To take the paramagnetic effect into account, we evaluate the Maki parameter using $\alpha_{\mathrm{M}}=\sqrt{2}H_{c2}^{\mathrm{orb}}/H_{\mathrm{P}}=0.88$, where $H_{c2}^{\mathrm{orb}}$ is the orbital limit for dirty superconductors determined by $\mu_{0}H_{c2}^{\mathrm{orb}}=-0.69\,T_{c}\,d(\mu_{0}H_{c2})/dT|_{T=T_{c}} = 11.0$ T and $H_{P}$ is the BCS Pauli paramagnetic limit, $\mu_{0}H_{p}=1.86\,T_{c}=17.7$ T. Using this value and $\lambda_{\mathrm{SO}}=1.1$, a theoretical curve accurately describes the measured $H_{c2}$ above $T=2$ K $\sim 0 .2\,T_{c}$. However, in contrast to the theoretical curves, the measured $H_{c2}$ increases linearly with decreasing temperature from $4~\mathrm{K}\sim 0.4\,T_{c}$ to $0.1~\mathrm{K}\sim 0.01\,T_{c}$. We note that the simple theoretical model for $H_{c2}$ of dirty two-gap superconductors \cite{gurev03} cannot reproduce linear-in-$T$ behavior of $H_{c2}$ observed in \sns. The anomalous temperature dependence of $H_{c2}$ observed in \sns\ is reminiscent of those for the iron based superconductors \ce{CaKFe4As4} \cite{brist20} and FeSe \cite{brist23}, in which a Fulde-Ferrell-Larkin-Ovchinnikov state, a pairing state formed between the Zeeman–split Fermi surfaces, is proposed to be stabilized. Although further experimental and theoretical studies, such as using a model taking account of Fermi surface shapes \cite{shiba06}, would be required, the linear-in-$T$ behavior of $H_{c2}$ down to $0.01\,T_{c}$ may suggest the realization of an exotic superconducting pairing state owing to spin-split Fermi surfaces due to asymmetric SOC in \sns.

Finally, we discuss peculiar features of \di\ in the superconducting state of \sns. We observe the prominent ZBCP at low temperatures, as shown in FIG.\ref{fig:ZBCPTH}(a)) and its unusual temperature and field dependence, as shown in FIGs.\ref{fig:ZBCPTH}(c) and (d). A ZBCP can appear in unconventional superconductors with nodes in the gap structure \cite{tanak95,yamas97} and topological superconductors hosting Majorana zero modes at the surface \cite{sasak11}. However, the extrinsic effect obscures potential intrinsic ZBCPs in our sample. The observed ZBCP at low magnetic fields can arise from the Josephson junction effect between superconducting domains \cite{lee71,shan03}, given that the superconducting volume fractions of our \sns\ single crystals are relatively small. The unusual enhancement of the ZBCP near $T_{c}$ and $H_{c2}$ can also be attributed to this effect. Indeed, similar enhancement due to the Josephson junction effect has been reported in the point contact spectroscopy of an \ce{La_{0.9}Ce_{0.1}CuO4} film \cite{he18b}. Further experimental efforts are required to clarify the presence of intrinsic ZBCPs in \sns.

\section{Summary}
In summary, we have studied the normal and superconducting states of \sns\ using soft-point-contact spectroscopy. Prominent asymmetric double peaks are observed in \di\ of the normal state. A phenomenological double Fano resonance model can explain the asymmetric double peaks. Our observations uncover the presence of flat energy bands and the hybridization between these flat energy and bulk bands in \sns. We also observe a dip feature in \di\ around $V_{\mathrm{bias}}=0$ in the normal state due to an opening of a pseudogap. The ratio of the gap amplitude to the gap-closing temperature for the hybridization gap, $2\Delta_{\mathrm{hyb}}(0)/k_{\mathrm{B}}T_{\mathrm{hyb}}$, agree with that for the pseudogap, $2\Delta_{\mathrm{PG}}(0)/k_{\mathrm{B}}T_{\mathrm{PG}}$, suggestive of a hybridization-driven pseudogap. In the superconducting state of \sns, a remarkable ZBCP emerges, characterized by unexpected temperature and magnetic field dependencies. However, the ZBCP likely originates from extrinsic effect. Furthermore, the out-of-plane upper critical field displays a linear-in-$T$ behavior at low temperatures down to $0.01T_{c}$, significantly diverging from the theoretical curves calculated by the WHH theory. This unusual behavior may suggest a possible exotic superconducting state associated with spin-split Fermi surfaces in \sns. Our findings suggest that \sns\ host topological surface flat bands arising from bulk nodal lines near its Fermi energy and the hybridization between the topological surface flat bands and bulk conduction bands causes the opening of the pseudogap.

\begin{acknowledgments}
The authors thank G. He, K. Jin, Y. Kasahara, and T. Shibauchi for useful discussions. This work was supported by an NSF Career DMR-1944975.
\end{acknowledgments}

\bibliographystyle{apsrev4-2_YN.bst}

\end{document}